# Polarimetric properties of asteroid 3200 Phaethon

L. F. Golubeva[1], D. I. Shestopalov[1], O. I. Kvaratskhelia[2]

[1]Shemakha Astrophysical Observatory, Azerbaijan Academy of Science, Shemakha AZ-5626, Azerbaijan
(lara_golubeva@mail.ru , shestopalov_d@mail.ru)
[2]Evgeni Kharadze Natiornal Astropihysical Observatory, Abastumani 0300, Georgia
(kvara_otar@mail.ru)

**Abstract**

The polarimetric observations of asteroid 3200 Phaethon, the target of international observation campaign, did not cover a proper phase angle interval to provide estimating all the attributes of the asteroid polarization curve. Based on present discrete observation data for Phaethon, its full polarimetric curves (i.e., the phase angle – polarization degree dependences) in BVRI bandpasses were reproduced. We used a simple approximate formula, which proved itself well in analyzing asteroid, comet and lunar polarimetric data (Shestopalov and Golubeva, 2015; 2017). The negative polarization branch of Phaethon was found to be approximately the same for various observation sets, ($P_{min} \approx -1.35\%$ at $\alpha_{min} \approx 10^o$ and the polarimetric slope $h \approx 0.22\%/^o$ at the inversion angle $\alpha_i \approx 21^o$), whereas the positive polarization branch ($P_{max} \approx 64\%$, $51\%$, $47\%$, and $43\%$ at $\alpha_{max} \approx 135^o$) depends on the observation date. We reexamined the geometric and Bond albedo of the asteroid (0.089 and 0.04, respectively) as well as the size of regolith particles (~ 400 μm) on its surface. We found also arguments in favor of the idea of extensive terrain in the asteroid northern hemisphere with highly altered surface originated seemingly in near-sun environment. In general the polarimetric properties of the asteroid correspond to a notion on surface structure as thermally altered regolith particles mixed with lager rock fragments like a coarse pebble.

## 1. Introduction

A rocky body about 6 km in diameter, called 3200 Phaethon, belongs to Apollo-type near Earth asteroids (NEAs) and due to its very elongated orbit approaches the Sun at a distance of about 0.14 AU, which is approximately 2 times less than the perihelion of Mercury. Phaethon is the largest body among potentially hazardous asteroids (PHAs) and near-Sun asteroids. In addition, Phaethon is probably a parent body of the Geminids meteor shower and can be an inactive cometary nucleus with a depleted stock of volatile fractions. It is clear, why the asteroid with such outstanding properties is a very attractive object for many-sided astrophysical observations, especially intensive during the periods of close proximity to the Earth (see Lazzarin et al., 2019; Taylor et al., 2019; Kim at al., 2018; Kinoshita et al., 2017; Fornasier et al., 2006; Krugly et al., 2002 in more detail).

The high-precision polarimetric observations of Phaethon at various wavelengths were also carried out by several teams of observers (Devogèle et al., 2018; Ito et al., 2018; Zheltobryukhov et al., 2018; Borisov et al., 2018) during its approaches to the Earth in 2016 and 2017. These investigations aimed to explore the phase-angle dependences of linear polarization degree in the widest possible range of the angles. But it so happened that the phase angle coverage in the datasets amounted to 33–117$^{o}$. As a result, the main attributes of the asteroid polarimetric curves (such as their extrema $P_{min}$, $P_{max}$, the value of inversion angle $α_i$, polarimetric slope $h$ at $α_i$) proved to be uncertain. As respects to the $P_{max}$, this occurs for the reason that Phaethon has unusual high polarization degree, and polarimetric curve reaches its maximum at the phase angles more than 120$^{o}$ (e.g., Devogèle et al., 2018). The absence of the quantitative characteristics of the asteroid phase-polarization dependences adversely affects the subsequent interpretation of polarimetric observations.

A goal of this paper is to retrieve all possible information from the available polarimetric measurements of Phaethon in order to improve our understanding of the asteroid surface properties. This is the interesting challenge, on our opinion, since the next closest approach of the asteroid to the Earth will be in 2050. At least for the next five years, the observational conditions will be unfavorable for the polarimetry of Phaethon in the range of maximum polarization degree. One would be expected that the full polarimetric curves of Phaethon itself and its surface units would be obtained during the space mission DESTINY+, which is planned to be launched in 2022, but the spacecraft is not planned to be equipped with polarimeter (Arai, 2017). In the next sections of the paper we state a correspondence between aforesaid polarimetric measurements of Phaethon and characteristics of its phase-polarization curves, test the validity of our extrapolations, and estimate the asteroid geometric and Bond albedos as well as the grain size on Phaethon's surface. Finally, we will discuss the findings.

**2. The phase-polarization curves of asteroid Phaethon**

The degree of the linear polarization of light scattered by planetary surface is commonly defined as follows:

$$P_r = (I_\perp - I_\parallel)/(I_\perp + I_\parallel). \tag{1}$$

Here $P_r$ is so-called signed polarization degree that can be both a negative value when the plane of polarization is parallel to scattering plane (in this case the intensities $I_\perp < I_\parallel$), and a positive value when the plane of polarization is perpendicular to scattering plane ($I_\perp > I_\parallel$). We recall that the scattering plane contains the incident and scattered rays, and the phase angle $\alpha$ is bounded by these rays.

If the phase angle lies between $0°$ and inversion angle $\alpha_i$, at which the polarization degree changes sign, then $P_r$ is found to be negative. The main characteristics of the negative branch of polarimetric curve are the position of minimum $P_{min}$ at $\alpha_{min}$ and polarimetric slope $h = dP_r(\alpha)/d\alpha$ at $\alpha = \alpha_i$. When $\alpha > \alpha_i$, the polarization degree becomes positive and reaches its maximum $P_{max}$ at $\alpha_{max}$.

The problem of Phaethon's polarimetric observations (Devogèle et al., 2018; Ito et al., 2018; Zheltobryukhov et al., 2018) is that the phase angle coverage constitutes 33–117°, so aforesaid characteristics of polarimetric curves cannot be directly derived from these datasets. Indeed, the inversion angle $\alpha_i$, a frontier between the positive and negative branches of polarimetric curves, must be substantially less than 33° since a sole observation of Phaethon in the vicinity of $\alpha_i$ performed by Fornasier et al. (2006) gives $P_r$ = 0.43% at $\alpha$ = 23°. Because of high polarization degree at $\alpha > 50°$, $P_r$ could reach its maximum at $\alpha \sim 130°$ in compliance with the estimates made by Devogèle et al. (2018) and Zheltobryukhov et al. (2018). Thus the first step in analyzing the asteroid polarimetric properties is approximation of the polarimetric curves of Phaethon by continuous functions according to the known set of discrete data.

For this task, one can apply the following simple empiric relations. One of them is a four-parameter ($b$, c1, c2, $\alpha_i$) trigonometric function

$$P(\alpha) = b \sin^{c1}(\alpha) \times \cos^{c2}(\alpha/2) \times \sin(\alpha - \alpha_i) \tag{2}$$

suggested by Lumme and Muinonen (1993). The second expression contains five free parameters ($m$, $n$, $l$, $h$, $\alpha_i$) and was found by Shestopalov (2004):

$$P(\alpha) = B\left(1 - e^{-m\alpha}\right)\left(1 - e^{-n(\alpha - \alpha_i)}\right)\left(1 - e^{-l(\alpha - \pi)}\right), \tag{3}$$

where

$$B = \frac{h}{n\left(1 - e^{-m\alpha_i}\right)\left(1 - e^{-l(\alpha_i - \pi)}\right)}.$$

We have previously investigated the properties of Eq. (2) and found that the best fit of the negative polarization branch of bright lunar craters by this equation is accompanied with overestimating the positive polarization degree. And vice versa, when the best fit of positive polarization branch takes place (for the same bright craters), the negative polarization branch is strongly deformed (Shestopalov and Golubeva, 2015). Obviously, the same effect works for all high-albedo targets, for instance E-type asteroids.

Zheltobryukhov et al. (2018) have tried to apply Eq. (2) to fit Phaethon's observations performed by Ito et al. (2018). But they could not find a reasonable fit to the polarimetric curve of this asteroid by Eq. (2). Therefore a fourth-degree polynomial was calculated to reproduce the observations.

Belskaya et al. (2017) have also employed Lumme-Muinonen's function for refining the asteroid classification based on polarimetric observations. However, the mean values of the polarimetric slope $h$, which were found in Belskaya et al. (2017) to characterize the optical types of asteroids, lead to the correlation equation between asteroid geometric albedo and the slope, which is strongly different from the analogous correlations found previously by other investigators (e.g., Cellino et al., 2015; Shestopalov and Golubeva, 2015; Lupishko, 2018). It is seems, approximating the polarimetric curves of asteroids by Eq. (2) results in an overestimation of the slope $h$. Thus, one can conclude that Lumme-Muinonen's relation does not possess the universality and can lead to misunderstandings, which are difficult to foresee.

In turn, Eq. (3) has proved to be a good tool for reproducing with a good accuracy the polarimetric curves not only of the main-belt asteroids at the phase angle less than $30^o$ but also lunar areas and lunar soils, near-Earth asteroids, comets, the Moon (Shestopalov, 2004; Shestopalov and Golubeva, 2014; 2015; 2017), and Mercury (this paper) over a wide range of phase angles. Therefore we apply Eq. (3) to calculate the best approximation to the phase–polarization dependences for Phaethon and other objects that are under consideration here.

Figure 1 represents the best-fitting phase curves of Phaethon in the different bandpasses according to the observations of Devogèle et al. (2018) and Ito et al. (2018). The data received by Fornasier et al. (2006) in V bandpass were also allowed for the reducing of uncertainty in the inversion angle position. The probable values of asteroid polarimetric parameters obtained from these polarimetric functions are listed in Table 1. A set of two observations of Phaethon at different phase angles obtained by Zheltobryukhov et al. (2018) is insufficient for the reconstruction of reliable polarimetric curve, but we will not disregard their observations and discuss potential consequences in Section 6.

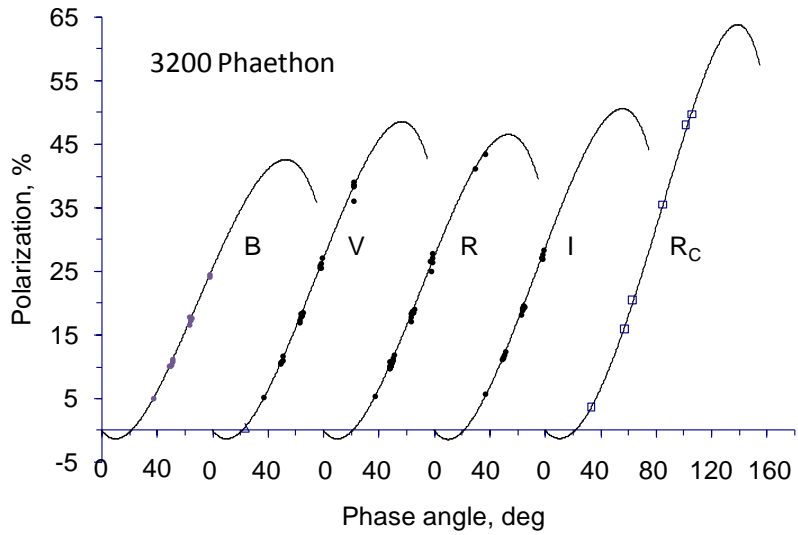

Fig. 1. The phase dependences of polarization measured at different wavelengths for asteroid 3200 Phaeton and their best fitting by Eq. (3). The effective wavelengths of BVRI and $R_c$ bandpasses are 435, 559, 685, 830, and 641 nm, respectively. Observation data were taken from Devogèle et al., 2018 (filled circles), Fornasier et al., 2006 (triangle), and Ito et al., 2018 (open squares). Error bars are comparable to the size of symbols on the plot. The curves are shifted from each other along the x-axis to avoid overlaps.

As is seen in Table1, the probable value of $P_{max}$ parameter for Phaethon in autumn 2016 is substantially higher than that in winter 2017. It supports the prognosis made by Zheltobryukhov et al. (2018), though there is a difference in numerical values obtained by these authors and us ($P_{max}$ = 54% and 63.9%, respectively). Incidentally, when using polynomial approximation, as Zheltobryukhov et al. have done, the shape of fitted curve (for instance, the value and position of its extrema) depends on the degree of polynomial, the choice itself which is not obvious of.

Unlike the positive polarization degree, the parameters of the negative polarization branch of Phaethon slightly vary with respect to each other. We will check below whether our polarimetric characterization of Phaethon is physically reasonable.

## 3. Asteroid Phaethon on the $P_{max}$–$\alpha_{max}$ plot

We have already used this diagram in order to be convinced in validity of our inference about the parameters which remain obscure for the main-belt asteroids, that is the value and position of

Table 1. The probable values of the main polarimetric parameters of asteroid Phaethon by observations in 2016 (Ito et al., 2018) and in 2017 (Devogèle et al., 2018).

| Date | Bandpass | $m$ (deg$^{-1}$) | $n$ (deg$^{-1}$) | $l$ (deg$^{-1}$) | $\alpha_i$ (deg) | $h$ (%×deg$^{-1}$) | $\Delta_h$ | $P_{min}$ (%) | $\Delta_P$ | $\alpha_{min}$ (deg) | $\Delta_{\alpha min}$ | $P_{max}$ (%) | $\alpha_{max}$ (deg) | $\Delta_{\alpha max}$ |
|---|---|---|---|---|---|---|---|---|---|---|---|---|---|---|
| 14-17/12/2017 | B | 0.0247 | 0.0 | -0.0288 | 21.50 | 0.218 | 0.010 | -1.34 | 0.30 | 10.1 | 1.3 | 42.6 | 132.8 | 0.3 |
| 14-19/12/2017 | V | 0.0238 | 0.0 | -0.0356 | 21.42 | 0.222 | 0.018 | -1.35 | 0.54 | 10.1 | 2.3 | 48.6 | 136.6 | 0.6 |
| 14-20/12/2017 | R | 0.0234 | 0.0 | -0.0298 | 21.24 | 0.225 | 0.012 | -1.35 | 0.35 | 10.0 | 1.5 | 46.6 | 133.6 | 0.4 |
| 14-17/12/2017 | I | 0.0240 | 0.0 | -0.0339 | 21.39 | 0.237 | 0.009 | -1.43 | 0.28 | 10.0 | 1.1 | 50.6 | 135.7 | 0.3 |
| 15/09-07/11/2016 | R$_C$ | 0.0145 | 0.0 | -0.0358 | 21.72 | 0.220 | 0.023 | -1.29 | 0.63 | 10.4 | 2.7 | 63.9 | 139 | 0.5 |

*Notes:*

The bandpasses are centered at the following wavelengths (Devogèle et al., 2018 and Ito et al., 2018): B = 435 nm, V = 559 nm, R = 685 nm, R$_C$ = 641 nm.

$\Delta_h$, $\Delta_P$, $\Delta_{\alpha max}$, $\Delta_{\alpha min} \approx \Delta_{\alpha i}$ are standard errors of the slope parameter $h$, the maximum degree of positive and negative polarization $P_{max}$ and $|P_{min}|$, their phase angles $\alpha_{max}$, $\alpha_{min}$, and the inversion angle $\alpha_i$ respectively. They were calculated by the same procedure that was applied in (Shestopalov and Golubeva, 2015).

polarization maximum (Shestopalov and Golubeva, 2015). In the mentioned paper we have found that linear correlation between the predicted values of $P_{max}$ and $\alpha_{max}$ in case of large E and S asteroids with high and moderate albedo is actually similar to that observed by Dollfus and Bowell (1971) for lunar areas. Such a coincidence can seemingly occur if the surfaces of the bodies are close in the composition and structure of upper light-scattering layer. This assumption is quite reasonable since the surfaces of these asteroids and the Moon are formed by finely dispersed regolith and consist of widespread mafic silicates such as olivines and pyroxenes. In turn, the high degree of positive polarization obviously distinguishes Phaethon from the asteroids with the known polarization properties. Thus, in order to expand the search of physical analog of Phaethon's surface, the lunar soil gathered during cosmic missions was also taken into consideration. To control our inferences about Phaethon's surface we also involved the polarimetric measurements of Mercury (Dollfus and Auriere, 1974), since it is the nearest planet to the Sun, the properties of the regolith of which are known quite well (e.g., Kiselev and Lupishko, 2004). In addition, the disk-integrated polarimetric measurements of Mercury were carried out in a wide range of phase angles that makes it possible to calculate the exact phase-polarization curve of the planet and evaluate all of the required polarimetric characteristics.

The laboratory observations of lunar samples were performed by Kvaratskhelia (1988) with the same polarimeter and software that were exploited for telescopic observations of the Moon. This neat solution ensured approximately equal measurement errors of 0.1 – 0.15% in polarization degree in astronomical and laboratory observations. Some lunar samples were separated by size of their particles, other remained "all-in-one-piece". Several polarimetric curves for the lunar soil samples with different size of particles are presented in Fig. 2. For comparison, the phase-polarization curve of Mercury is also shown here.

Relations between the maximum polarization degree $P_{max}$ and the corresponding phase angle $\alpha_{max}$ for the aforesaid objects and Phaethon are shown in Fig. 3a. It appears to be two $P_{max}$–$\alpha_{max}$ sequences, and both sequences being invariant for wavelengths in the visible spectral range. The first of them (solid line) is formed by planetary bodies with a regolith covering and corresponds to the linear equation

$$\alpha_{max} = 98.1° + 0.65 P_{max}$$

with correlation coefficient $r = 0.87$. The second sequence (dashed line) is composed of the samples of lunar soils delivered to the Earth by Luna 16, 20, 24 probes and Apollo 16 spacecraft from continental and maria regions of the Moon. The asteroid Phaethon seems to belong to this sequence, which is described by the following equation:

$$\alpha_{max} = 86.13 P_{max}^{0.118}$$

with correlation coefficient $r = 0.95$.

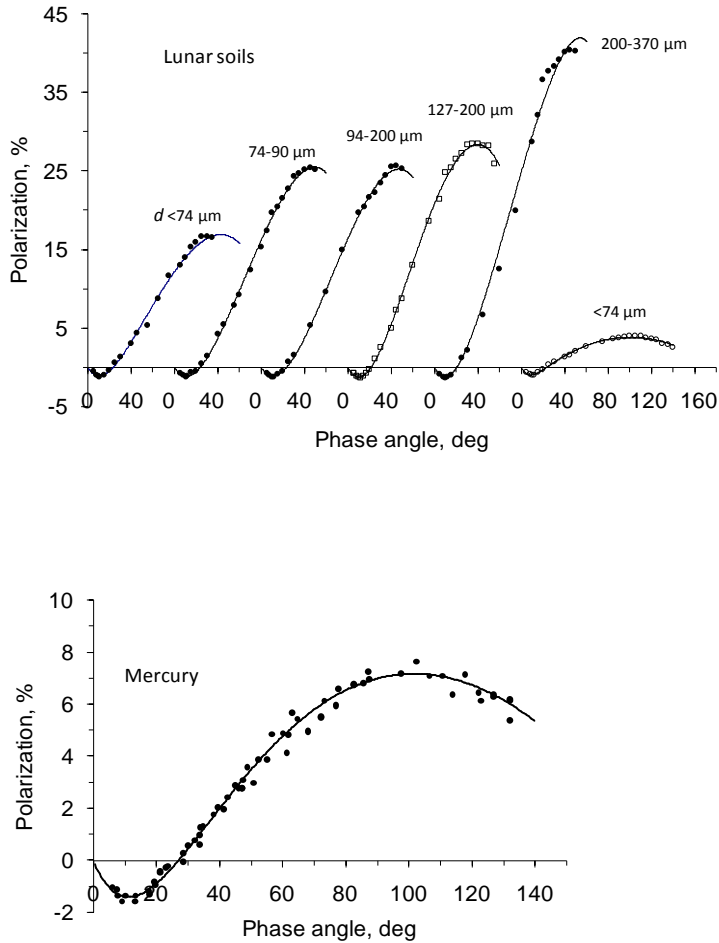

Fig 2. (*top panel*) The best approximation of the phase-polarization curves of lunar soil samples obtained with Eq. (3). The samples were delivered by the *Luna 24* probe, *Luna 16* probe, and *Apollo 16* spacecraft and were subsequently separated by the grain size *d*. These respective curves are marked by filled points, open squares, and open points. Initial polarimetric data in the visible range ($\lambda = 550$ nm) are taken from the polarimetric catalog by Kvaratskhelia (1988). The errors in polarization measurements were found to be 0.15%. The errors of approximation are ~0.2-0.4% in polarization degree and ~1° in the phase angle. The error bars is comparable to the size of symbols on the plot. The curves are shifted from each other along the abscissa to avoid overlaps. (*bottom panel*) The best fit with Eq. (3) to the phase–polarization dependence observed for Mercury by Dollfus and Auriere (1974) at $\lambda = 580$ nm. Mercury has the following polarimetric characteristics: $P_{min} = -1.4$ %; $\alpha_{min} = 11.1°$; $P_{max} = 7.18$ %; $\alpha_{max} = 101.7°$; $h = 0.142$ %/deg.

We paid attention to the following circumstance (Fig. 3b). If lunar samples remained undivided, then their $P_{max}$ and $\alpha_{max}$ values lie nearby the linear sequence. But the separation of samples by grain size leads to an increase of both polarimetric parameters, which tend to be greater, the greater the size of particles. In Fig. 3b, the separated dust-free fraction of lunar soil from Mare Crisium (Luna 24 probe) with particle size of 200–370 μm terminates on the right the second nonlinear sequence and is similar in the $P_{max}$, $\alpha_{max}$ parameters to those of asteroid Phaethon. Thus we can suppose that Phaethon's surface might be composed of a substance resembling coarse sand with a particle size of several hundred microns.

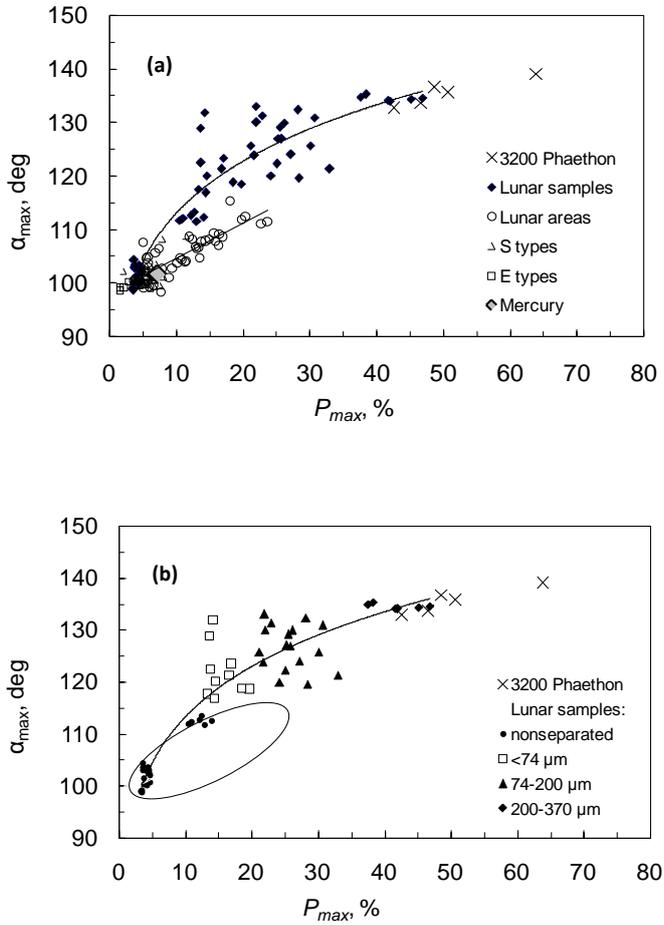

Fig. 3. (a) Relations between the polarization maximum $P_{max}$ and the phase angle $\alpha_{max}$, at which the polarization maximum is achieved, for the lunar areas, lunar samples, S-, E-type asteroids, Mercury, and asteroid Phaeton are presented. Dashed line shows an interrelation between the parameters of the lunar samples; solid line is the same for the lunar areas. The initial data for the construction of the plot were taken from Kvaratskhelia (1988), Dollfus and Auriere (1974) and Shestopalov and Golubeva (2015). (b) The same as the previous plot but the grain size of lunar soil samples are specified. Ellipse bounds the position of planetary bodies covered by regolith layer.

Appealing to the data for Mercury, we hoped to find at least some correspondence between constitutions of light-scattering layer on Phaethon and this nearest-to-Sun planet. But Mercury belongs, seemingly, to the first (linear) sequence though the space weathering level of its surface higher than on the Moon and, a fortiori, main-belt asteroids. The polarimetric properties and consequently the structure and grain size of Mercurian regolith proved to be similar to the lunar one.

**4. Geometric and Bond albedos of asteroid Phaethon**

Several estimates of the visual geometric albedo, $p_V$, of Phaethon are known to day. Based on data from IRAS satellite, Tedesco et al. (2002) were the first to find Phaethon's geometric albedo to be 0.107±0.011. Hanuš et al. (2016) developed a new thermodynamical model of the asteroid body and, taking IRAS and Spitzer thermal data, derived $p_V$ = 0.122±0.008. Radar observations of Phaethon during the 2017 apparition demonstrated a roughly spherical body with average diameter of 6.2 km (Taylor et al., 2019). This asteroid size estimation leads to $p_V$ = 0.068 proceeding from the absolute magnitude in the V band, i.e., $H$ = 14.6 (Minor Planet Center Database, 2019; https://www.minorplanetcenter.net/iau/MPCORB/NEAp08.txt).

Now we estimate the visual geometric albedo of Phaethon $p_V$ deduced from polarimetric measurements (or polarimetric albedo in short). For that we use the predicted parameters $P_{min}$ and $h$ in the V bandpass from Table 1 and the empirically established relationships

$$\log p_V = C_1 \log h + C_2$$

and

$$\log p_V = C_3 \log |P_{min}| + C_4 .$$

It should be noted that we apply decimal logarithms in these formulae and hereinafter. The above equations are a consequence of Umov's effect: the linear polarization degree of light scattered from rough surface inversely correlates with the albedo of the surface. Estimating the $C_i$ coefficients is the subject of ongoing research aimed at refining these indices. The values of the coefficients taken from the publications of recent years (Shestopalov and Golubeva, 2015; Cellino et al., 2015; Lupishko, 2018) are given in Table 2. The albedo data sources shown in the Table illustrate the fact that the above-mentioned researchers used different data sets; therefore their studies do not duplicate each other. Although the sample sizes and approximative procedures were also different in the referred works, the equation coefficients placed in columns of Table 2 are equivalent values from the standpoint of statistics. Such a behavior of $C_i$ coefficients has been previously discussed (Cellino et al., 1999; Shestopalov and Golubeva, 2015). It should be noted Cellino et al. (2016) analyzed the photometric observations of asteroid Vesta from Dawn probe together with the groundbased polarimetric measurements of the asteroid and found good agreement between the slope-albedo relation derived from Vesta's data and the same relation obtained by Cellino et al. (2015) for asteroids.

Table 2. The geometric albedo of Phaethon estimated from polarimetry.

| $\log p_V = C_1 \log h + C_2$ | | $\log p_V = C_3 \log |P_{min}| + C_4$ | | Albedo data | Phaethon's polarimetric albedo (this paper) | | Ref.[a] |
|---|---|---|---|---|---|---|---|
| $C_1$ | $C_2$ | $C_3$ | $C_4$ | | $p_V(h)$ | $p_V(|P_{min}|)$ | |
| -1.078±0.069 | -1.725±0.064 | -1.286±0.097 | -0.821±0.019 | AKARI | 0.095±0.004 | 0.103±0.007 | [1] |
| -1.087±0.084 | -1.793±0.075 | -1.337±0.128 | -0.893±0.026 | WISE | 0.083±0.004 | 0.086±0.008 | [1] |
| -1.111±0.031 | -1.781±0.025 | -1.419±0.034 | -0.918±0.006 | Stellar occultations | 0.088±0.001 | 0.079±0.002 | [2] |
| -1.016±0.010 | -1.719±0.012 | -1.331±0.015 | -0.882±0.016 | AKARY, WISE, Stellar occultations | 0.088±0.001 | 0.088±0.004 | [3] |

*Notes:*
[a] the $p_V$–$h$ and $p_V$–$P_{min}$ calibrations for the last 5 years: [1] Shestopalov and Golubeva (2015), [2] Cellino et al. (2015), [3] Lupishko (2018).

As we cannot give preference to either of the calibrations of asteroid polarimetric albedo, all eight estimates of Phaethon's geometric albedo are offered in Table 2. These values constitute also uniform dataset since the differences between any pairs of $p_V$ in this set are within two standard errors. Therefore, a simple average of these eight estimates equal to 0.089±0.007 is the relevant estimate of Phaethon's polarimetric albedo. The derived value lies in the interval between aforesaid estimates of radiometric and radar albedo of the asteroid. In addition we can conclude that our characterization of the negative polarization branch of Phaethon is quite pertinent.

To estimate Bond albedo $A_{sp}$ for Phaethon in the V bandpass we took the equation

$$\log A_{sp} = (-1.1 \pm 0.1) \log h - (2.1 \pm 0.1)$$

found in (Shestopalov and Golubeva, 2011). The predicted value $h$ from Table 1 gives $A_{sp} \approx$ 0.04±0.01. We assume that $A_{sp}$ of Phaethon can be even lower, say, 0.03–0.035 to provide the slope parameter *G* in *GH* photometric system to be similar to that for low-albedo asteroids (Warner et al., 2009).

To determine the optical type of the asteroid with more confidence, we applied the following technique. Knowing $p_V$ and the characteristics of the negative polarization branch of Phaethon, as well as the average values of the same parameters for the various asteroid types (i.e., the coordinates of their centers in the space of optical characteristics), we can calculate the distance between the tested asteroid and the center of the given asteroid type (*t*):

$$D(type) = \sqrt{\begin{array}{l}((p_V^{Ph} - p_V^t)/k_1)^2 + ((P_{\min}^{Ph} - P_{\min}^t)/k_2)^2 + ((\alpha_{\min}^{Ph} - \alpha_{\min}^t)/k_3)^2 \\ + ((\alpha_i^{Ph} - \alpha_i^t)/k_4)^2 + ((h^{Ph} - h^t)/k_5)^2\end{array}}, \quad (4)$$

where each weighting factor $k_i$ is specified in the same units as the respective optical parameter.

When calculating the distance *D* we are based on the new results obtained due to refining asteroid taxonomy realized by Belskaya et al. (2017). The weighting factors $k_i$ in Eq. (4) should be chosen in such a way, that the terms in the above formula did not prevail one over the other. For $k_i$, we chose the maximum values from the list of the average values of the polarimetric parameters of the asteroids of various optical types (Table 3 in the referred work). Like that $k_1 = 0.51$, $k_2 = 1.9\%$, $k_3 = 13°$, $k_4 = 28°$. As we have mentioned above, there is a problem in determining polarimetric slope *h* in the work of Belskaya et al. (2017), therefore the last summand in Eq. (4) was omitted. Now we can use Eq. (4) since the optical characteristics of Phaethon have been found here, and the same average values for asteroid types were derived by Belskaya et al. (2017).

So, the smaller the parameter *D*, the closer to each other the optical characteristics of the tested asteroid and asteroids of the given optical type. Figure 4 shows the distribution of the distances *D* in relation to the asteroid types. The uncertainty in *D* is caused by the errors in optical parameters inherent to asteroid types and Phaethon itself. Calculations show that the choice of the weighting factors $k_i$ has a little effect on the shape of the distribution. As follows from Fig. 4, the smallest *D* = 0.151 corresponds to P type, and little farther from Phaeton are situated C and B types ($D$ = 0.170 and 0.171, respectively). The differences between these three values lye within their errors, and so it is difficult to give preference to any one of the derived estimates.

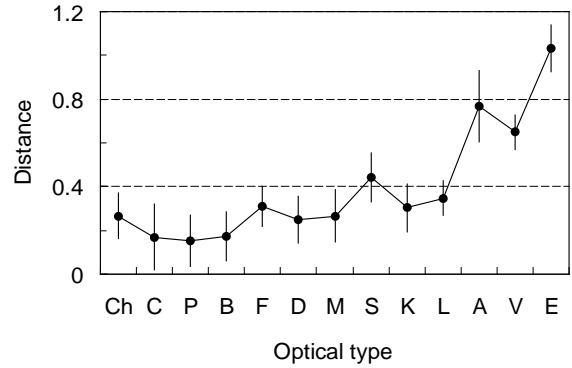

Fig. 4. The distance $D$ between Phaethon and the coordinates of the center of optical types in the space of $p_V$, $P_{min}$, $\alpha_{min}$, and $\alpha_i$ variables. Uncertainty in $D$ is produced by the errors in the optical parameters that determinate the asteroid types and Phaethon itself.

The belonging of Phaethon to B type was determined by Binzel et al. (2001), though the asteroid was initially classified by Tholen (1985) as a member of F class (see Devogèle et al., 2018 in more detail). Later, the variability of spectral slope was found by several investigators in the wavelength region shortward ~0.55 μm (Lazzarin et al., 2019 and references therein) resulting seemingly from the asteroid surface heterogeneity (see also Section 7 here). Under these circumstances, B- or F-type classification of Phaethon and our taxonomic estimate should be considered as the first approximation, but without a doubt the asteroid belongs to the C-complex, which combines the C/G/B/F/P asteroid types. It would be expected that multicolor photometry of Phaethon with high spatial resolution from aboard a space probe DESTINIY+ (Arai, 2017) will be able to specify the optical taxonomy of asteroid surface units.

## 5. Asteroid Phaethon on albedo–$P_{max}$ plot

Traditionally, albedo and $P_{max}$ parameters are believed to be linked to the particle size of particulate surface. The question in relation to Phaethon was studied by Ito et al. (2018) and Zheltobryukhov et al. (2018) rather on a qualitative than quantitative level. Such a situation occurred because of the uncertainty in the position of Phaethon on the albedo – $P_{max}$ diagram, where the measurements of lunar soil, terrestrial rocks, meteorites and other objects are collected together (Dollfus and Titulaer, 1971; Dollfus and Auriere, 1974; Geake and Dollfus, 1986; Ito et al., 2018). The $A$–$P_{max}$ plane in logarithmic coordinates is divided into two half-planes, one of which is occupied by the measurement data, and the other is empty. A dividing line is given by the equation derived from telescopic observations of small areas on lunar surface (Dollfus and Bowell, 1971):

$\log A = (-0.724 \pm 0.005)\log P_{max} - (0.36 \pm 0.02)$.

As was later shown, this relation for lunar and the most likely asteroid surfaces weakly depends on the wavelength, at least in the visible region of the spectrum (Shkuratov and Opanasenko, 1992; Shestopalov and Golubeva, 2015; Zheltobryukhov et al., 2018). A body of similar laboratory measurements was fulfilled for lunar fines samples and powdered terrestrial rocks; at that all the data were found to lie above the lunar line. Apparently such a behavior of $A$ and $P_{max}$ parameters reflects the fact that the specific type of roughness of planetary regoliths is very difficult to reproduce in laboratory.

We use the $A$–$P_{max}$ diagram constructed by Geake and Dollfus (1986) where the generalization of the data obtained by the authors and their colleagues was made as follows. The regions occupied by the lunar and powdered terrestrial samples are shown by straight regression lines. They are almost parallel, but their position on the diagram depends on the grain size of the powdered and separated samples. When the grain size increases, then the intercept coefficient of regression equations between $A$ and $P_{max}$ is also increases. As a result, these regression lines are offset with respect to each other. The farthest region from the lunar dividing line is occupied by solid terrestrial rocks in their natural state with dusty and dust-free surfaces. On this diagram (Fig. 5), we showed position of Phaethon and also Mercury to verify the correctness of our actions.

The measured albedo $A$ for the objects in the referred telescopic and laboratory datasets were reduced to the phase angle of $5^{\circ}$ (Geake and Dollfus, 1986), whereas geometric albedo $p_V$ of Phaethon and Mercury corresponds to zero phase angle. Obviously it is incorrect to compare these values. To put Phaethon and Mercury correctly in their places on the $A$–$P_{max}$ diagram we used the correlation between $A$ and polarimetric slope $h$ derived by Geake and Dollfus (1986) for the lunar fines (<50 μm in particle size) and the powdered terrestrial and meteorite samples (with the particle sizes of 50–340 μm):

$$\log A = (-1.026 \pm 0.038)\log h - (1.873 \pm 0.039).$$

It is important in the given context that the coefficients of the relationship are slightly depend on the grain size and texture of light scattering surfaces (Geake and Dollfus, 1986) and also on wavelength at least in the visible spectral range (Shestopalov and Golubeva, 2016). So, for Phaethon and Mercury we have $A = 0.062$ and $0.099$ that corresponds their polarimetric slopes $h = 0.222$ and $0.142$, respectively. In Fig. 5, Mercury is located near the lunar line that apparently indicates the similar structural properties of lunar and Mercurian regoliths. The $A$, $P_{max}$ parameters of Phaethon fall into the region occupied by the nonseparated natural terrestrial rocks with solid surfaces sometimes powdered with dust. Therefore, we can imagine Phaethon's surface as a mixture of pebble-stone and coarse sand with grains of about 300 microns in size. We note also the mutual agreement between the findings received in this and the third section.

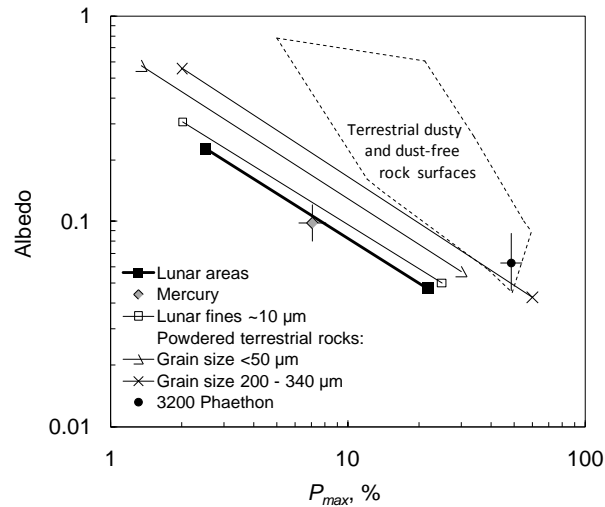

Fig. 5. Albedo $A$ against the maximum polarization degree $P_{max}$ for the lunar and terrestrial samples as provided by Geake and Dollfus (1986). Interrelations between these optical parameters are shown for the lunar fines samples (the grain size is about 10 μm), the powdered terrestrial rocks (the grain sizes are <50 μm and ~200–340 μm), and the natural dusty and dust-free rock surfaces. Regression line for the lunar areas (telescopic observations) according to Dollfus and Bowell (1971) and the position of Mercury and Phaethon are also shown.

**6. Estimating the particle size on Phaethon's surface**

In lunar literature, one can find several approaches intended for remote sensing the particle size of lunar regolith (Dollfus, 1998 and references therein). However the authors warn that the derived empiric relations between the median grain size of lunar regolith on one hand and albedo and the maximum polarization degree on the other hand should be applied with caution to other airless Solar System bodies (Shkuratov and Opanasenko, 1992; Dollfus, 1998). Such a constraint arises owing to specific cohesive fluffy structure of lunar regolith, as a product of long-time impact disruption of mother rocks together with thermal metamorphism resulting in agglutination and vitrification of surface particles. In a sense, Mercurian surface microstructure reminds the lunar regolith (e.g., Kiselev and Lupishko, 2004) but it is exactly not the case of Phaethon's surface (Figs. 3, 5).

Considering close location of the lunar samples and Phaethon in Fig. 3, we decided to take as a basis the polarimetric measurements of the separated lunar soil samples (Kvaratskhelia, 1988) for estimating the particle size on the surface of Phaethon. Previously, of course, the discrete phase-polarization measurements of the lunar samples were represented by the approximating curves with the help of Eq. (3) to calculate as accurately as possible all the polarimetric parameters of the studied samples (see Fig. 2 as example). Next, the polarimetric parameter should be chosen to characterize the particle size on Phaethon's surface. As is known from asteroid physics, the minimum and maximum polarization degree and polarimetric slope as well depend on the albedo of light-scattering surface. In general terms, albedo is the complicated function of the composition, structure, and texture of the surface. Albedo at the given surface composition depends mainly on the particle size of the light scattering layer and, to a lesser extent, on soil roughness and porosity. Thus it is possible only by experience to select the necessary polarimetric parameter(s) capable of limiting the albedo effect on the measured value of target parameter.

Recently, we have found the values of $P_{max}$ for a set of main-belt asteroids with the known negative polarization branch; the predicted values have been also tested for appropriateness (Shestopalov and Golubeva, 2015). It is easily to check by the data in Table 1 in the referred paper that $P_{max}/P_{min}^2$ ratio for the studied asteroids does not depend on their albedos (unlike $P_{max}$ or $P_{min}$ themselves), and consequently could be sensitive to other characteristics of surface.

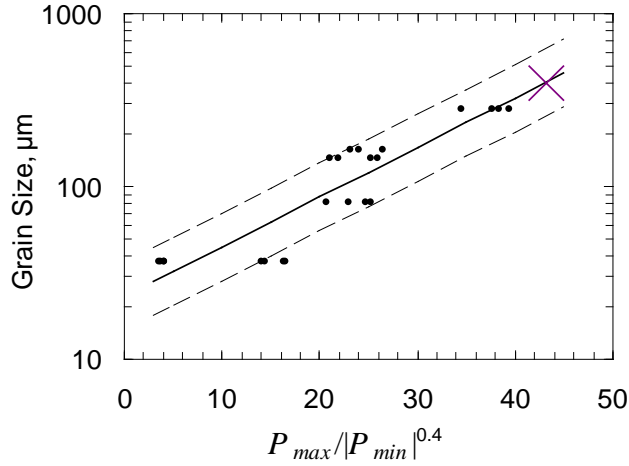

Fig. 6. The grain size of separated lunar samples comparing to their $P_{max}/|P_{min}|^{0.4}$ ratio in the visible spectral range. The data scatter area is bounded by the dashed lines. The location of Phaethon on the plot is marked with a cross. Initial data were taken from Kvaratskhelia (1988) for separated lunar samples and Devogèle et al. (2018) for asteroid Phaethon.

Since relatively large asteroids of our set are also covered by fine-grained regolith, we searched for interrelation between $P_{max}/|P_{min}|^a$ ratio and the grain size of the separated lunar samples. A free parameter $a$ was varied from 2 to 0.05 and was found to be close to 0.4 in accordance with the maximum coefficient of correlation between the particle size and the above ratio. The wavelength of measurements slightly affects the computation result owing to the fact that the particle size of the samples is much larger than the wavelength in the visible spectrum. A solid line in Fig. 6 presents the direct regression relations obtained for the lunar samples by the white-light observations and at wavelengths of 533.5, 550, 641.5 nm. Assuming that the particle size distribution in the separated lunar samples is uniform, the size of the particles is determined by their median diameter *d*. The regression line corresponds to the following equation:

$$\ln d = (0.066 \pm 0.007) \times P_{max}/|P_{min}|^{0.4} + (3.14 \pm 0.16), \qquad (5)$$

with the correlation coefficient $r = 0.9$. Since $P_{max}/|P_{min}|^{0.4} = 43.1 \pm 7$ for Phaethon in the visible region, then the averaged grain size on its surface amounts to 400 μm (Fig. 6). According to the error of the above polarimetric ratio the probable limits of the grain size variation on the asteroid surface were found to be 630–250 μm. In concordance with the results of the previous sections we would give preference the upper limit of the granulometric estimate. By the way, the observations of Phaethon performed by Ito et al. (2018) in the autumn of 2016 lead to the ratio $P_{max}/|P_{min}|^{0.4} = 57.7$. This value corresponds to $d \sim 1000$ μm if, of course, Eq. (5) remains valid at such large values of the polarimetric ratio.

A few words about the polarimetric observations of Phaethon by Zheltobryukhov et al. (2018) would be appropriate. Their white-light observations were performed at two phase angles of 57.9° and 73.2° on 16 and 17 December 2017. The polarization degree at the latter phase angle appreciably exceeded those reported by Ito et al. (2018) and Devogèle et al. (2018). Based on the fact that polarimetric observations at $\alpha = 73.2°$ corresponded to a minimum of the asteroid brightness variation, the authors put forward a suggestion on a large-scale heterogeneity of Phaethon's surface. This viewpoint seems to be supported by radar observations of the asteroids by Taylor et al. (2019) and the rotationally resolved spectral observations by Lazzarin et al. (2019). Both observations were performed in the same time period as the polarimetric observations on December 2017. The former showed roughly 600-m radar-dark spot near Northern Pole. The latter indicated the variability of asteroid optical spectrum in the wavelength range of 0.33–0.58 μm at the same viewing geometry. Besides, Borisov et al. (2018) based on the own observations of the asteroid on 15 December 2018 found a variation of the linear polarization degree according to Phaethon's rotational phase.

In turn, we have calculated the probable polarimetric curve of the asteroid using three observational results (Zheltobryukhov et al., 2018; Fornasier et al., 2006) and, as normal, three roots of Eq. (3) at $\alpha = 0°$, $180°$, and $\alpha_i = 21.9°$. These six points are the minimum data set that allows the curve to be calculated. Our simulation of Phaethon's polarimetric properties for the given observations results in $P_{max} \approx 78\%$ at $\alpha_{max} \approx 136°$ that are unusual for NEAs and main-belt asteroids. Of course, the uncertainty of the forecast of these values increases because of the small number of observations. Penttilä et al. (2005) investigated the problem of predicting polarimetric values for main-belt asteroids and comets and found the probable errors for the polarization maximum and its phase angle: $\Delta P_{max}/P_{max} \sim \pm 0.3$ and $\Delta \alpha_{max} \sim \pm 10°$. However, we decided not to ignore the observations by Zheltobryukhov et al. (2018) as we have found a suitable pattern in the polarimetric catalog by Kvaratskhelia (1988). The maximum polarization degree for the slagged lunar breccia (the Luna-16 collection) reaches even $88.7\% \pm 1\%$ at $\alpha_{max} = 136.4°$ (Fig. 7). Seemingly, polarimetric, photometric, spectral, and radar observations sense together a spacious low-albedo region on Phaethon's surface. It appears also that the idea of Phaethon's surface burned by solar insolation (Ohtsuka et al., 2009; Hanuš et al., 2016) receives a further support

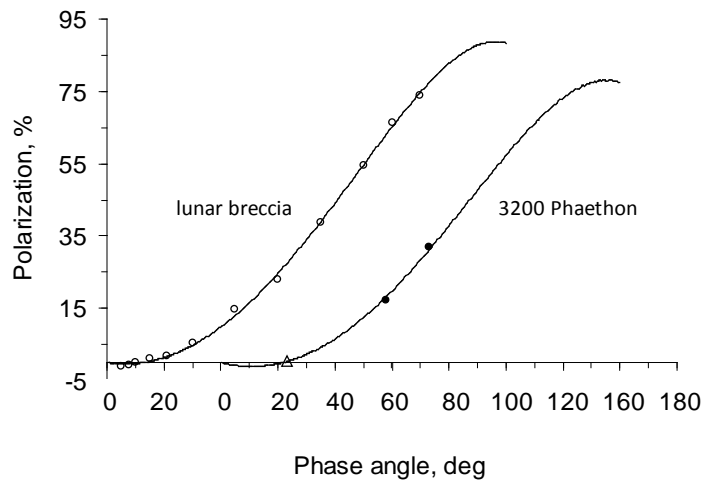

Fig. 7. The phase-polarization curves of lunar breccia with the slagged surface (polarimetric measurements by Kvaratskhelia, 1988) and asteroid Phaethon according to polarimetric observations by Zheltobryukhov et al. (2018) performed on 16 and 17 December 2017. Polarimetric observation at the phase angle of 23° (triangle) by Fornasier et al. (2006) are also used when computing the polarimetric curve of Phaethon. The error bars of measurements are comparable to the size of symbols on the plot.

## 7. Discussion and conclusion

The peculiarity of the polarimetric observations of Phaethon consists in the fact that the measurements were carried out under such conditions that do not allow all the parameters of the asteroid polarimetric curves to be directly determined from the observational data. These polarimetric parameters of the asteroid were derived from its phase-dependent polarization curves fitted to the observation data with Eq. (3). Since the obtained polarimetric parameters are the extrapolated values it was necessary to check their acceptability. Simultaneously this procedure realizes quantitative analysis of the polarimetric properties of Phaethon. The quantitative estimates are found to be quite relevant to determine geometric and Bond albedos, and the size of the regolith particles. In addition, we adduce the evidence to back up the arguments for extensive terrain near the asteroid Northern Pole with the highly altered surface originated seemingly in near-sun environment.

Due to the specific orientation of rotation axis, the northern hemisphere of Phaethon undergoes a thermal shock during several weeks around its perihelion (Ohtsuka et al., 2009). Phaethon moves in osculating orbit, its perihelion distance in accordance with Hanuš et al. (2016) was the lowest about 2 kyr ago (~0.126 AU as opposed to 0.14 AU in the current epoch). At that time the subsolar equilibrium surface temperature in the asteroid northern hemisphere, especially in the high Northern latitudes, could reach ~ 1220 K (we use Phaeton's thermal model from Ohtsuka et al., 2009 and our estimate of the asteroid Bond albedo from Section 4). Such a high-temperature heating of the surface, when the asteroid rotation axis almost coincides with the direction of the Sun, can be accompanied by the intensive thermal metamorphism of extended near-pole region (Ohtsuka et al., 2009). So we can expect degassing and dehydration of surface rocks, perhaps their decomposition as well as puffing, vitrification and agglutination of surface particles. Such factors of space weathering as solar wind and micro-meteoroid bombardment, although they also increase during perihelion passage, cannot soften thermally induced degradation of Phaethon's surface. Solar wind spattering does not affect surface composition and structure but alters the spectral and brightness properties of asteroid surface. Microparticle bombardment at typical relative velocity of ~5 km/s in the main asteroid belt results generally in crushing the agglomerates formed near perihelion as a result of sintering the non-hardened ductile surface particles. At collision speeds of more than 20 km/s at distances of less than 1 AU from the Sun, the melting of surface particles can form impact breccias with the slagged surface, which are common in mature lunar soil.

In the past, about 500 yr ago and again ~ 4 kyr ago, Phaethon's southern hemisphere was turned towards the Sun at perihelion (Hanuš et al., 2016). All the ensuing consequences have already been discussed above in the case of the Northern hemisphere of the asteroid. It is interesting, the rotationally resolved spectral observations of Phaethon in 2007 showed the color variability of its southern hemisphere depending on the rotation angle of the asteroid: from the B-type blue colors to the C-type neutral or slightly reddish colors (Kinoshita et al., 2017). The authors estimated the area ratio occupied by B-type and C-type colors in Phaethon's southern hemisphere to be approximately 80/20. Seemingly, the surface inhomogeneity near Southern Pole of Phaethon is also discovered.

Owing to weak gravity, the body can persistently lose the smallest particles of the surface, which can be removed by light pressure from the terminator zone. Some fraction of particles of intermediate size can be lost due to shaking the surface in collisions with larger projectiles. Apparently the typical asteroid surface relief is thermally altered coarse-grained regolith mixed with larger rock fragments that substantially corresponds to the polarimetric properties of the asteroid.

The inferences of our work will be probably tested in the course of space mission DESTINY+ to the asteroid. In particular, observations in situ with high spatial resolution will seemingly be able to solve the problem of optical taxonomy of asteroid surface. However, the full phase polarization curves of 3200 Phaethon itself and, more importantly, its surface units will remain empirically undetermined for many years.

**Acknowledgements**

We express our appreciation to all the observers whose remarkable observation data formed the basis of our work. We are deeply appreciative to E. V. Petrova for her careful review and thoughtful discussion.